\documentclass{aa}
\usepackage{graphicx}
\usepackage{txfonts}
\usepackage{natbib}
\begin{document}
\authorrunning {J. X. Cheng et al}
\titlerunning{statistical study of RHESSI hard X-ray spikes}
   \title{Solar flare hard X-ray spikes observed by  RHESSI: a statistical study}
   \author{J. X. Cheng \inst{1,2} \and J. Qiu  \inst{3} \and  M. D. Ding   \inst{1,4}
          \and  H. Wang\inst{5}}
    \institute{School of Astronomy $\&$
Space Science, Nanjing University, Nanjing 210093, China  \\
              \email{chengjx@shao.ac.cn}
              \and Shanghai Astronomical Observatory, Chinese
              Academy of Sciences, Shanghai 200030, China
              \and Department of Physics, Montana State University, Bozeman MT
59717-3840, U.S.A.     \and Key Laboratory of Modern Astronomy and
Astrophysics (Ministry of Education), Nanjing University, Nanjing
210093, China \and
        New Jersey Institute of Technology, 323 Martin Luther Kind Blvd., Newark, NJ 07102, U.S.A.}
   \date{Received 2012 January 12; accepted 2012 September 14}
  \abstract
   {Hard X-ray (HXR) spikes refer to fine time structures on timescales of seconds
to milliseconds in high-energy HXR  emission profiles during solar
flare eruptions.}
   {We present a preliminary
statistical investigation of temporal and spectral properties of
HXR spikes. }
 {\upshape Using a three-sigma
spike selection rule, we detected  184  spikes in  94  out of 322
flares with significant counts at given photon energies, which
were detected from demodulated HXR light curves obtained by the
\itshape Reuven Ramaty High Energy Solar Spectroscopic Imager
\upshape (RHESSI). About one fifth of these spikes are also
detected at photon energies higher than 100 keV.}
{The statistical properties of the spikes are as follows. (1) HXR
spikes are produced in both impulsive flares and long-duration
flares with nearly the same occurrence rates. Ninety percent of
the spikes occur during the rise phase of the flares, and about
70\% occur around the peak times of the flares. (2) The time
durations of the spikes vary from 0.2 to 2 s, with the mean being
1.0 s, which is not dependent on photon energies. The spikes
exhibit symmetric time profiles with no significant difference
between rise and decay times. (3) Among the most energetic spikes,
nearly all of them have harder count spectra than their underlying
slow-varying components. There is also a weak indication that
spikes exhibiting time lags in high-energy emissions tend to have
harder spectra than spikes with time lags in low-energy
emissions.}
   {}


   \maketitle
%

\section{Introduction}
Solar flare emission at sub-second timescales was reported in hard
X-ray (HXR) observations by satellite-borne HXR spectrometers in
the 70s and 80s \citep{van74,van76,hoyng76,jager78, kiplinger83,
kiplinger84,kiplinger89}. Examining HXR flares observed by the
\itshape Solar Maximum Mission \upshape (SMM) with time
resolutions of 128 ms and 10 ms, \citet{kiplinger83} found that 53
out of nearly 3000 flares  produce several hundred fast spikes
with durations as short as 45 ms. These energetic flare bursts on
short timescales are believed to be nonthermal in nature, and
their temporal and spectral properties place constraints on the
physical nature of the source.

 Several physical mechanisms are proposed to produce HXR spikes as rapid and short-lasting enhancement over the
slow-varying underlying emission. Magnetic reconnection is
believed to provide the means of converting energy stored in the
magnetic fields into thermal and kinetic energies
\citep[e.g.,][]{car64,Stu66,hira74,kopp76,priest00,asch04}.
Particles may be accelerated to high energies at the reconnection
site in the corona and then deposited at the chromosphere emitting
HXR radiation by bremsstrahlung. It is known that along the length
of the pre-connection current sheet, a tearing mode instability
may occur to trigger reconnection and form magnetic islands
\citep{furth63,Stu66}. Formation of the magnetic islands by
tearing instability and subsequent interactions between these
magnetic islands, namely the dynamic magnetic reconnection
\citep[and references therein]{kliem00}, are considered to account
for fast variations in nonthermal emissions \citep[and references
therein]{aschwanden02}. \citet{kaufmann96} also proposed that in
the magnetically complex solar active regions, one would expect to
have multiple, primeval explosive compact synchrotron sources
flashing at different times, building up what is usually described
as the onset of the impulsive phase of the bursts. Different
spikes may indeed originate at various sites, as suggested by one
mm-wave observation \citep{correia95}. The primeval compact
sources might originate from a number of plasma instabilities,
such as in twisted magnetic fields, and magnetic flux networks
\citep{sturrock81,sturrock84}. Another proposed mechanism to
produce fine structures in HXR and microwaves concerns nonthermal
electron injections. Electron beams become unstable when fast
particles of a nonthermal electron distribution outpace the slower
ones from the thermal or suprathermal tail during their
propagation along magnetic fields lines. These unstable electron
beams cause discrete electron injections and finally contribute to
spiky HXR emissions.

The kinematics of energized electrons and the energy dependence of
their collisional interactions in high-density plasmas in solar
flares can be probed by accurate time delay measurements between
 HXR emission at different energies. There is a variety of the time
delay mechanisms operating in solar flares that can be used as a
diagnostics of physical parameters. In the thick-target model
\citep{brown71,fisher85, emslie85,mac84,mar89} for HXR emission in
solar flares, electron acceleration is assumed to occur in flaring
loops at coronal heights, while HXR bremsstrahlung emission is
produced in the chromosphere. Under this assumption, the velocity
spectrum of the accelerated electrons causes time-of-flight
differences that are expected to result in a delay of the lower
energy HXR emission with respect to that in higher energies
\citep{asch95,brown98}. A different case is that the HXR emission
in higher energies lags behind the emission in lower energies
\citep{aschwanden97,qiu04}.  Two basic models have been proposed
for the latter:  the trap-plus-precipitation model \citep{mel76,
vil82,bespalov87} and the second-step acceleration model
\citep{bai79}. In the trap model, the lag of the high-energy
emission occurs because the collisional timescale increases with
the particle energy. This delays the escape from the trap and the
subsequent precipitation and, thus, the resulting thick-target HXR
emission. In the second-step acceleration model, a second-stage
process is invoked to accelerate super-thermal electrons to higher
energies. Both of these models produce an energy delay with a sign
opposite to that caused by the time-of-flight effect. In
particular, whether the low-energy emission lags or the
high-energy emission lags  depends on the relative importance of
these two mechanisms. Involved in different mechanisms during the
flare eruptions, we study the meaningful energy-dependent time
delay that occurs in the spikes with  high temporal and spectral
resolving observations.

The \itshape Reuven Ramaty High Energy Solar Spectroscopic Imager
\upshape \citep[RHESSI,][]{lin02} has observed several tens of
thousands of flares with unprecedented temporal and spectral
resolution since its launch in early 2002. It uses a set of nine
rotating modulation collimators, each consisting of a high
spectral resolution germanium detector. As the spacecraft rotates,
the grids transmit a rapidly time-modulated fraction of the
incident flux. To identify fine time structures in flare HXR
emissions observed by RHESSI, a demodulation algorithm was
recently developed to remove the modulation pattern from the HXR
light curves. The demodulation code was applied to some flares
that were found to produce fast-varying spikes evident in HXR
emissions of over 100 keV \citep[][hereafter, Paper I]{qiu12}. Ten
spikes analyzed in Paper I exhibit a sharp rise and decay with a
duration of less than 1 s, with the designated instrument
resolution of 125 ms. Energy-dependent time lags are found in some
spikes, which are consistent with the time-of-flight of
precipitating electrons estimated from imaging observations. All
these spikes show a harder spectrum than the underlying
components, indicating that the spectral properties are closely
related to timescales.

These  ten  spikes analyzed in Paper I are produced in a few
randomly selected flares. It is important to conduct a systematic
search for fast spikes using existing databases. In the present
study, we examine all flares observed by RHESSI in 2002, and apply
the demodulation code to flares with  more than 100 data counts in
12--25 keV to identify HXR spikes in a few photon energy bands. We
then conduct a statistical study of spike-productive flares and
the temporal and spectral properties of HXR spikes. The paper is
organized as follows. Observations and data analysis are given in
\S2. Properties of HXR spikes are presented in \S3, followed by
discussions and conclusions in \S4.
\section{Observations and data analysis}

\subsection{Selection of HXR spikes}

Using the demodulation algorithm as described in Paper I, we
examined several  RHESSI observations of flare bursts for HXR
spikes, and analyzed their temporal and spectral properties. Our
database includes all  RHESSI  HXR flares in the year 2002.
According to the RHESSI  flare list, 5302 flares were observed in
2002. Among them, we selected relatively strong events with peak
count rates in 12--25 keV greater than 100 counts s$^{-1}$ and
significant emission in the photon energy range greater than 50
keV. This information was directly acquired from the RHESSI flare
list. Only 322 flares satisfied these two pre-analysis selection
criteria and were analyzed with the demodulation code. These are
our final sample for the statistical study.

\begin{figure}
   \includegraphics[width=9cm]{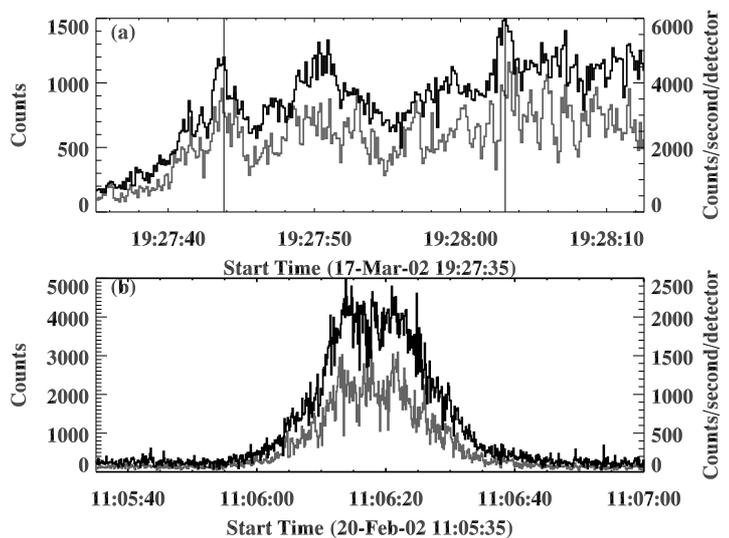}
      \caption{Demodulated
RHESSI  HXR light curves (dark) in comparison with summed raw data
counts (gray). The light curves are acquired at 25--100 keV.  The
demodulated light curves correspond to the count rate y-axis and
the summed raw data are related to the count y-axis.
 The top panel shows an event on 2002 March 17 that
exhibits fast-varying spikes. The bottom panel shows an event on
2002 February 20 without significant fast-varying spikes. The
scales of the raw and demodulated light curves are indicated in
the left and right y-axis, respectively. }
 \label{FigVibStab}
 \end{figure}

We applied the demodulation algorithm to obtain high temporal
resolution light curves with a designated cadence of 125 ms.
Figure 1 shows examples of high-candence flare HXR light curves at
25--100 keV that are demodulated from the raw data. A large number
of fluctuations are present in the raw data set, most of which,
however, were removed in the demodulated light curves. Clearly,
not all flares produce sufficiently significant HXR spikes to be
picked out by the selection rule. Figure 1a shows an event with
evident fast-varying structures, the spike at 19:27:43 UT, while
for the event in Figure 1b no outstanding fast varying spikes can
be recognized in the demodulated light curves.  One cannot tell
whether a flare produces fast-varying spikes from undemodulated
raw data.

To identify spikes, we subtracted from demodulated light curves
$I(t)$ a slow-varying component $I_{slow}(t)$, which is obtained
by box smoothing $I(t)$ over a smoothing window $w_{smt}$.
$w_{smt}$ is taken to be between 4 and 8 s, or 32 to 64 bins. We
then obtained the residual intensity as
$I_{r}(t)=I(t)-I_{slow}(t)$. We defined a spike when the residual
intensity, $I_r(t)$, in the 25--100 keV band was $\geq$
$n_{sig}\sigma$ for three consecutive time bins, where $\sigma$ is
the standard deviation of the residual intensity profile, and
$n_{sig} = 3, 4, 5$. We used two ways to derive $\sigma$. First,
we derived a fixed value of $\sigma$ independent of flare
evolution, namely, $\sigma \equiv \sigma_0$, where $\sigma_0$ is
the standard deviation of $I_r(t)$ for the entire duration of the
flare, which is usually 2--3 minutes in our data
 (Figure 2a). Second, to minimize the effect of
statistical noises associated with instantaneous photon counts
received by the detector during the evolution of the flare, we
also used a time-dependent $\sigma(t)$ derived in a running box
$w_{sig}$  (Figure 2b-c). Derivation of $I_{slow}(t)$, $I_{r}(t)$
and $\sigma$ and selection of HXR spikes by these algorithms is
illustrated in Figure 2 using the event on 2002 August 3 as an
example. The left panels in the figure show the demodulated light
curve at 25--100 keV, superposed with the slow component
$I_{slow}$ derived with the three methods. The residuals $I_{r}$
 and $\sigma$ levels (smooth line) are also plotted at the bottom of each panel.
 The right panels illustrate the times (x-axis) and energy bands
 (y-axis: see figure caption) whose residuals exceeded
 3$\sigma$ level.
Evidently, the different choices of $\sigma$ by different
combinations of $w_{smt}$ and $w_{sig}$ result in different
populations of spike-productive flares, which will be discussed
below. We remark that the selection rules likely give a
conservative lower limit of the spike population because of the
requirement of three consecutive time bins.
\begin{figure}
   \includegraphics[width=9cm]{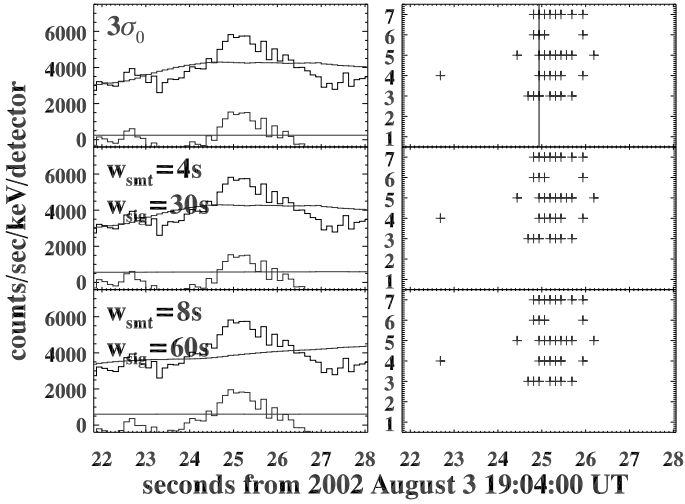}
      \caption{Spike detection algorithm using the three methods (see
text) for the event on 2002 August 3. Left panels show the
demodulated light curves  at 25--100 keV, the slow-varying
component (smooth) determined by the three methods, and the
residuals (bottom of each panel). The smoothed lines (bottom of
each panel) show the sigma levels using the three methods. Right
panels show the times (x-axis) and energies (y-axis) of
significant residuals recognized by the algorithms. The numbers 1
to 7 along the y-axis indicate energy bands of 10--15, 15--25,
25--40, 40--60, 60--100, 100--300, and 25--100 keV, respectively.
The vertical line indicates the most pronounced spikes  of the
diagrams. }
         \label{FigVibStab}
   \end{figure}

\begin{figure}
\includegraphics[width=9.cm]{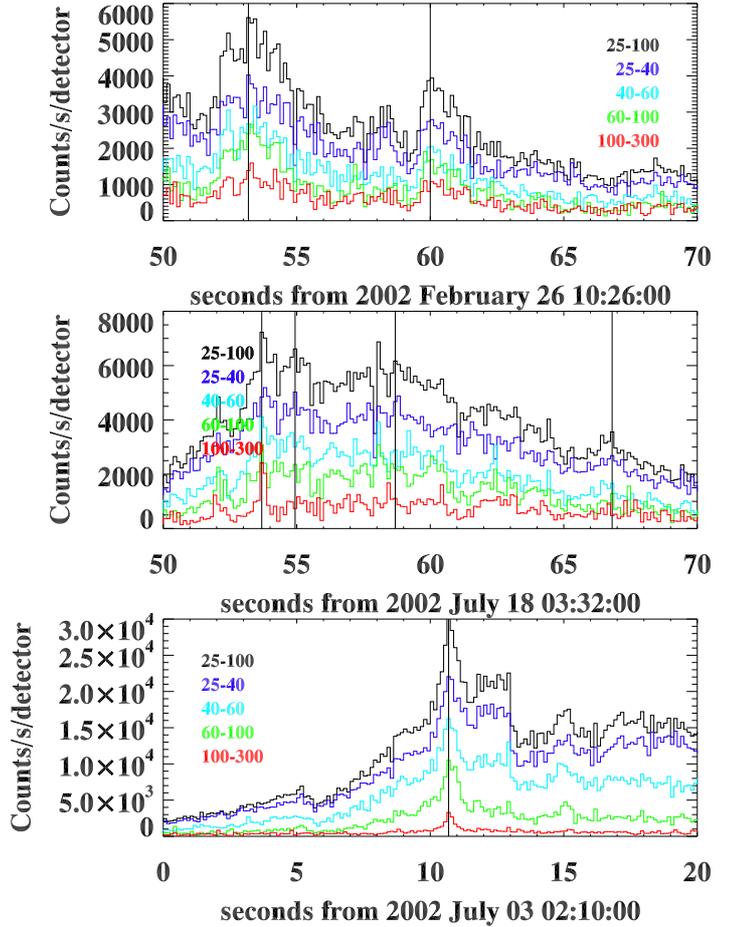}
\caption{Demodulated RHESSI HXR light curves at varying energies
for three flares. The vertical lines indicate the spike times. The
y-axis gives the units of the 25--100 keV light curves, and light
curves at other energies are scaled arbitrarily.}
\end{figure}

 \begin{figure}
   \includegraphics[width=9cm]{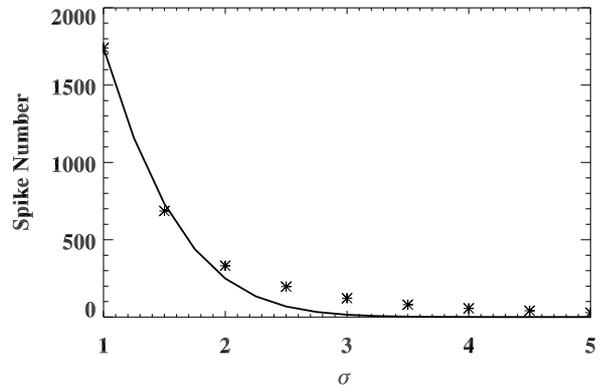}
      \caption{Number of detected spikes varies as $\sigma$ level increases.
 The asterisks and the line represent observational and  Poisson distributions, respectively.              }
         \label{FigVibStab}
   \end{figure}

We applied the selection rules to the 322 samples. Table 1 gives
the number of spikes and the number of spike-productive flares
using different combinations of $w_{smt}$ and $w_{sig}$, including
the fixed $\sigma_0$ criterion, and with $n_{sig} = 3$. Depending
on how $\sigma$ is defined, the selection rule of 3$\sigma$ and
three consecutive time bins yields 40 to 90 spike-produtive
flares, each producing one or two spikes, out of the total of over
5302 flares observed by RHESSI in 2002. This productivity is in
general comparable with the result by \cite{kiplinger83}, who,
also by applying the 3$\sigma$ and consecutive bin criterion,
detected 53 spike-productive flares out of 3000 flares observed by
 SMM. That is to say, fast-varying spikes are detected in 10--20\%
of flares, though observed by different instruments with different
methods.
\begin{table}
\caption{Spike/flare number variations with $\sigma$ defined by
different combinations of $w_{smt}$ and $w_{sig}$ and with
$n_{sig} = 3$. } \label{YSOtable}
\begin{tabular}{c c c c}
\hline\hline \ \ \ \ \ \ \ \ \ \ &   \ \ \ \ \ w$_{smt}$ = 4s \ \
\ \ \ & \ \ \ \ \ w$_{smt}$ = 6s \ \ \ \ \
&  \ \ \ \ \  w$_{smt}$ = 8s \ \ \ \ \   \\
\hline
  3$\sigma$$_{0}$ &  184/94  & -  & - \\
\hline
 w$_{sig}$=60s &  70/57 &  91/74  &  \ \ \ 103/80\ \ \\
\hline
w$_{sig}$=45s &  67/54  & 76/60 & \ \ \ 91/72\ \ \\
\hline
w$_{sig}$=30s &  47/39  &  57/46  & \ \ \ 65/51 \ \ \\
\hline
w$_{sig}$=15s &  28/25  &  27/23 &  \ \ \ 31/26 \ \ \\
\hline\hline
\end{tabular}
\end{table}

 Figure 3 shows the demodulated light curves of three
flares at varying energies from 25 to 300 keV during the flare
eruptions. The designated time bin is 0.125 s. These events all
exhibit fast-varying structures on timescale of 1 s or less at
most of the energy bins above 25 keV. Signals up to 100 keV are
significant and are likely real features. Some of these spikes are
also visible in 100--300 keV, though less significant because of
reduced counts level.

Using the fixed 3$\sigma_0$ yields 184 spikes in 94 flares. At
5$\sigma_0$, there are 54 spikes in 40 flares. At face value,
these numbers are significantly greater than the number of spikes
that could be produced by statistical noises if the noise
distribution followed a simple Poisson (Figure 4). In reality, the
noise distribution is poorly defined, and the noise dependence on
instantaneous photon counts is likely much stronger than a
Poisson. Therefore, we performed a negative test to identify
``negative" spikes by setting $n_{sig} = -3, -4, -5$ and using the
same rule of three consecutive bins. The results of the negative
test for a few representative selections are listed in Table 2.
Evidently, the negative occurrence is nearly one half of the
positive occurrence for the fixed $\sigma_0$. Even though one
cannot ascribe the negative occurrence entirely to the statistical
negative because of the way the residual signals are produced, it
is conservative to say that about one half of the positive
occurrence probably reflects real spike signals.

When the time-varying $\sigma$ is employed, understandably, the
number of spikes steadily decreases with a smaller $w_{sig}$ or
$w_{smt}$ (Table 1). Table 2 clearly shows that the negative
occurrence with the time-varying $\sigma$ drops to only 10--20\%
of the positive occurrence. Therefore, most spikes picked out with
the time-varying $\sigma$ approach may be considered as real
spikes free from statistical noises. We also note that 2/3 of the
spike population via the time-varying $\sigma$ approach are
included  in the population from the fixed 3$\sigma_0$ selection
rule.

\begin{table}
\caption{Spike/flare number variations with different
n$_{sig}$.}             
\label{table:1}      
\begin{tabular}{c c c c c c c}        
\hline\hline                 
n$_{sig}$ & 3 & -3 & 4 & -4 & 5 & -5 \\    
\hline                        
   $\sigma$$_{0}$ & 184/94&101/48 & 94/59  & 50/25  & 54/40  & 26/14\\      
   \hline
   w$_{sig}$=60s  \\
   \ \ & 103/80 & 10/9  &  36/30  & 4/4  & 14/10  & 3/3 \\
w$_{smt}$=8s\\
\hline
   w$_{sig}$=30s  \\
   \ \
 & 47/39 &  10/9  &  12/10 &  2/2 & 4/4 & 0/0\\
 w$_{smt}$=4s\\
\hline\hline                                  
\end{tabular}
\end{table}

The selection with the time-varying $\sigma$ does not only change
the total number of spikes but would, probably, change the
distribution of these spikes during the evolution of the flare.
For example, the fixed $\sigma_0$ regardless of the evolution of
the flare would misrepresent the noises at different evolution
stages. $\sigma_0$ tends to be smaller than the real standard
deviation determined by statistical counts during the flare
maximum but larger than the standard deviation during the
rise-and-decay phases of the flare when the counts are lower than
at the peak. Therefore, the constant $\sigma_0$ rule might include
noisy signals at the peak of the flare, while overlooking real
signals during the rise-and-decay phase. On the other hand, by
using the time-varying $\sigma$, the sample is partly biased
against high counts, or the peak phase of the flare.

Because we do not know the true noise distribution, we examined
and compared properties of spikes and associated flares selected
with the following different rules: fixed $\sigma_0$, time-varying
$\sigma$ with $w_{smt} = 8$ s and $w_{sig} = 60$ s, and
time-varying $\sigma$ with $w_{smt} = 4$ s and $w_{sig} = 30$ s.
The spikes were selected with $n_{sig} = 3$. Specifically, we
illustrated the distribution of the events with respect to the
flare magnitude, flare duration, and spike occurrence time, and
examine statistical properties of the spikes in terms of their
evolution and energetics.
\subsection{Event distribution}
We compared spike-productive flares with flares that do not
produce fast-varying spikes to examine whether they are
statistically different populations in terms of flare magnitude
and duration.
\begin{figure}
   \includegraphics[width=8cm]{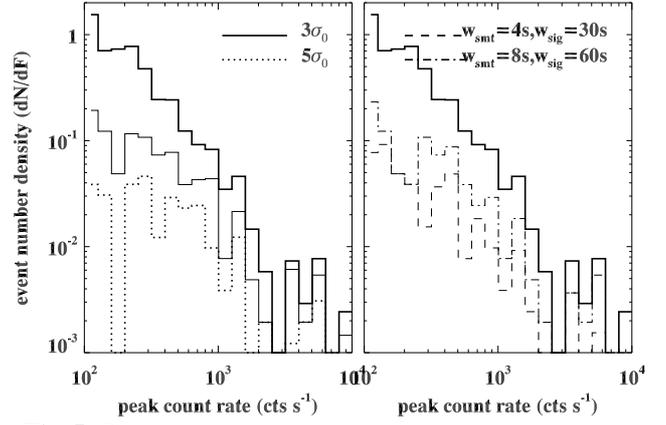}
      \caption{Peak
count rate distributions at various criteria denoted in the
figure. The upper thick solid line indicates our 322-flare sample.
The profile is obviously  steeper than all  other profiles.}
      \label{FigVibStab}
   \end{figure}
Figure 5 shows the peak count rate distributions of all  322
flares in the sample with a peak count rate exceeding 100, and the
spike-associated flares from four different selection rules:
5$\sigma_0$, 3$\sigma_0$, 3$\sigma$ at $w_{smt} = 8$ s and
$w_{sig} = 60$ s, and 3$\sigma$ at $w_{smt} = 4$ s and $w_{sig} =
30$ s. Figure 5 shows that the distribution for the entire sample
of 322 flares is steeper than that for the spike-associated
flares, suggesting that more intensive flares have in general a
greater chance to produce spikes. There is no notable difference
in the distributions of the four spike-productive flare
populations, suggesting that the majority of selected spikes are
not dominated by noise that grows with photon counts.

We also examined whether impulsive flares, compared with gradual
events, are more productive in fast-varying spikes. Figure 6
displays the rise time distributions for spike-productive flares
in comparison with that of all 322 flares. Here we define the
flare rise time as the time difference between the flare peak time
and the start time, both referring to emission in 12--25 keV
provided by the RHESSI flare list. No evident difference is found
between these distributions, suggesting that HXR spikes can occur
in both impulsive and gradual events.

\begin{figure}
   \includegraphics[width=8.5cm]{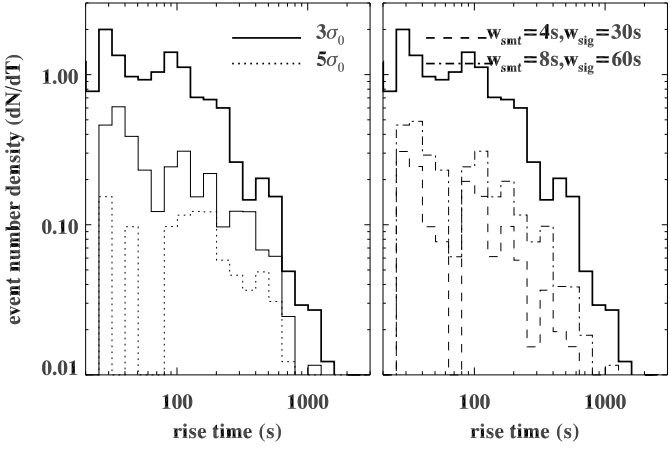}
      \caption{Same as Fig. 3, but for the
rise time of flares.}
         \label{FigVibStab}
   \end{figure}

For the spike-productive flares, we also studied at which time the
spikes occur with respect to the evolution stage of the flare
bursts. For convenience, we defined a normalized spike occurrence
time $\tau_{spk}$, which refers to the time difference between the
spike peak time and the flare start time normalized to the flare
rise time. Therefore, if $\tau_{spk} < 1$, the studied spike
should occur in the rise phase of the flare; if $\tau_{spk} > 1$,
the spike occurs during the decay phase. Figure 7 illustrates the
occurrence time distributions for the spike populations from
varying selection rules. For all populations, nearly all spikes
occur before or during the peak of the flare. Clearly, the fixed
$\sigma_0$ selection yields a large percentage, up to 70\%, of
spikes around the peak times ($\tau_{spk}$ = 0.8--1.2) of the
associated flares. With ever smaller $w_{sig}$ and $w_{smt}$, the
fraction of spikes during the flare peak is reduced as the
$\sigma$ around the flare peak time increases, while there are
more spikes detected during the low-count rise phase. At this
stage of research, as the noises of the demodulated flare light
curves are not well understood and cannot be independently
determined, we did not readily give a preference to a certain
selection rule. Therefore, all these spike populations from
different selection rules are analyzed in the following study.

\begin{figure}
   \includegraphics[width=9.5cm]{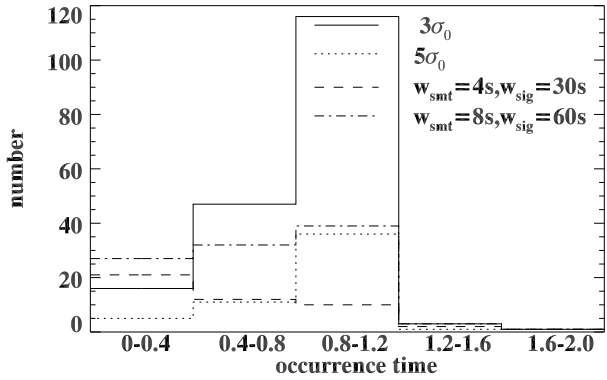}
      \caption{Occurrence time $\tau_{spk}$
 distribution at various criteria as denoted in the figure. See \S2.2 for details.}
         \label{FigVibStab}
   \end{figure}

Fast-varying spikes were first selected using integrated counts in
25--100 keV. We then demodulated the light curves of these events
in photon energy ranges of 25--40, 40--60, 60--100, and 100--300
keV, and searched for the highest energy in which spikes are
detected. We denote this energy band by $\epsilon_{hi}$. Figure 8
displays the spike distribution with respect to $\epsilon_{hi}$.
The majority of spikes discovered in 25--100 keV can still be seen
in 40--60 and 60--100 keV. Indeed, for each population, the peak
of the distribution is at 60--100 keV. Nearly 20\% of spikes can
be detected  as high as 100--300 keV. This result suggests that
fast-varying spikes are most probably nonthermal in nature.

\begin{figure}
   \includegraphics[width=9.5cm]{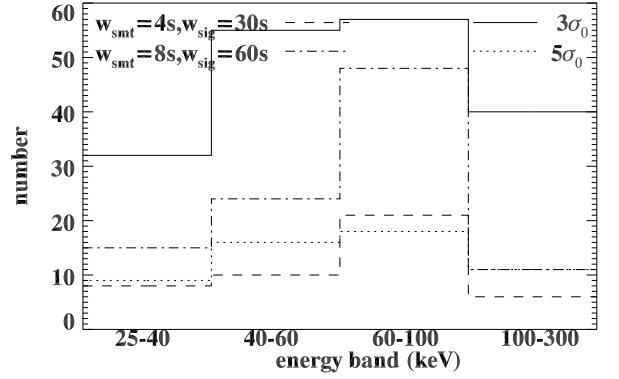}
      \caption{$\epsilon_{hi}$
distribution at various criteria. See \S2.2 for details.}
         \label{FigVibStab}
   \end{figure}

All these statistical results indicate that fast-varying spikes
are small-scale energetic events produced during the most
energetic stage of flares. In particular, they tend to occur
during the rise phase of intensive flares. On the other hand,
impulsive and gradual flares have an equal chance to produce
fast-varying spikes.

In different phases of flare eruption, energy release, and  energy
transport, the relevant physical processes are quite different.
Theoretically, energy release occurs preferentially during the
flare rise phase. This seems to support the result of different
productivities of spikes in different flare phases.

\section{Properties of HXR spikes}
HXR spikes at shortest timescales are believed to reflect single
energy release events, such as by magnetic reconnection and
particle acceleration. We examined the energy-dependent temporal
properties of these HXR spikes to gain insight into the nature of
the energy release. For the detected spikes, we estimated the
spike duration, rise and decay times, the count spectral index,
and the energy-dependent time delays and compared them  with large
flares and previous studies.

To obtain a statistical result of the temporal and spectral
properties of HXR spikes, we applied a simplified analysis to all
spikes. A more rigorous analysis was also conducted on a dozen
prominent spikes out of the selected samples (Paper I), which
yields results consistent with the statistical results in the
present paper.

\subsection{Duration, rise, and decay times}

The energy-dependent duration of a spike, $\tau (\epsilon)$, is
estimated to be the time difference between the e-slope start and
end times of the spike. We determined the rise and decay times
when the spike net intensity, i.e., the intensity with the
interpolated underlying component subtracted, drops to $e^{-1}$ of
the peak intensity in both the rise and decay phases. As displayed
in Figure 9, $\tau (\epsilon)$ ranges from 0.25 to 2.0 s, or 2 to
16 time bins, with an average being  0.98$\pm$0.47 s in 25--100
keV for all  HXR spikes. The same analysis was applied to light
curves in different energy bands, resulting in a mean duration of
 1.08$\pm$0.39  s, without significant difference in
different energies. Therefore, our studies show that the spike
duration is nearly independent of photon energies.

\begin{figure}
   \includegraphics[width=9.5cm]{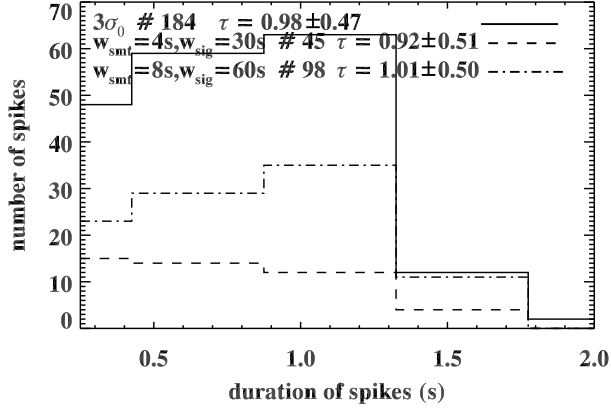}
      \caption{Duration $\tau (\epsilon)$ distributions
at various criteria. The average duration is about 0.9--1.0 s in
25--100 keV. see \S3.1 for details.}
         \label{FigVibStab}
   \end{figure}

HXR spikes appear to have quite a symmetric sharp rise and decay.
The mean e-slope rise time of all spikes is 0.52$\pm$0.26
 s, and the mean decay time is  0.50$\pm$0.25
 s. The result indicates no significant difference
between the rise and decay times. This is somewhat different from
the general scenario of flares, i.e., HXR emission rapidly
increases in the impulsive phase, and then decreases relatively
gradually in the decay phase.

\subsection{Spectral index}

\begin{figure}
   \includegraphics[width=9.cm]{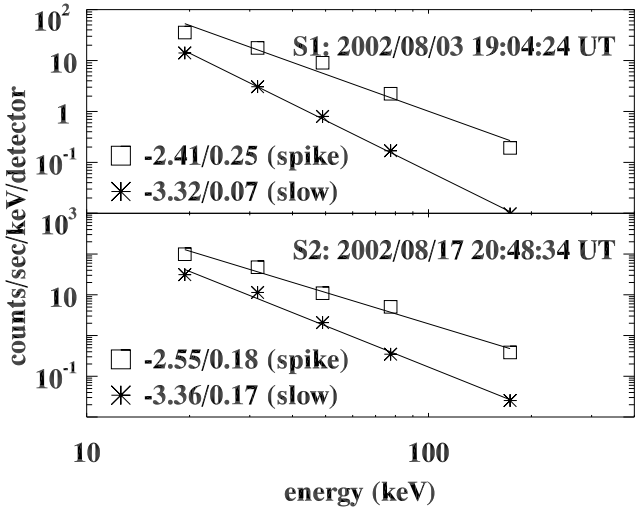}
      \caption{In each panel, the symbols and solid lines show examples of HXR count spectra of integrated spike flux during the spike time
in comparison with the count rates spectra of underlying
components. Solid lines show the least-squares fit of the spectra
to a power-law distribution, and the exponents and their
uncertainties from the fit are noted in the figures. The y-axis
gives the units of the spike intensities, and the intensities of
the underlying component are divided by a number for the clarity
of display. S1 is the example we discussed in Figure 2. }
         \label{FigVibStab}
   \end{figure}

To study the spike energetics, we fit the spike count spectrum at
$\ge$15 keV to a power-law distribution
$\frac{dI(\epsilon)}{d\epsilon}$ $\thicksim$
$\epsilon$$^{-\Gamma}$ to obtain the power-law index $\Gamma$,
 as indicated by the straight lines in Figure
10. Only data in photon energies greater than 15 keV were used to
avoid the contamination of thermal bremsstrahlung in lower
energies. Among these spikes, reliable least-squares fitting
results can be made for only a fraction of spikes. The spectral
fitting was applied to the net spike counts integrated over its
duration. The net counts of the spike were derived by subtracting
the underlying component, which is a mean between the pre-spike
and post-spike counts, also integrated over the duration of the
spike. As a comparison, we also fit the count spectrum of the
underlying component.  Figure 10 gives two examples of integrated
spike flux with respect to photon energy, i.e., the spike counts
spectrum. Also plotted are the directly measured count rate
spectra of the underlying components adjacent to the spikes.
 Figure 10 shows the comparison of spectral indices between the
spikes and underlying components. Note that $\Gamma$ is the index
of uncalibrated count spectrum, which is somewhat different from
the index of HXR photon spectrum or nonthermal electron spectrum.
However, it is still a meaningful parameter when comparing the
energetics of the spike relative to the underlying component.
Figure 11 shows that nearly  all spikes have a lower spectral
index than their underlying components by more than a
$\sigma_{\Gamma}$, the gross error in the fitting.
 Because we are aware of possible thermal contribution to photons
below 25 keV, we also fit the data from 25 keV for comparison. The
fitting result is nearly the same as that including the 15--25 keV
counts.
   This statistical result is consistent with the
more sophisticated case study in Paper I, indicating that most of
the spikes are more energetic, with a harder nonthermal spectrum
by about 0.5 in count spectrum index than the underlying
components

\begin{figure}
   \includegraphics[width=8.5cm]{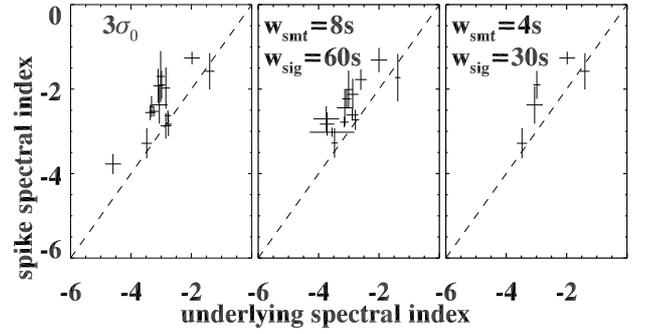}
      \caption{Scatter plot of spectral
indices of the spikes versus that of the underlying components
 using data from 15 keV. }
         \label{FigVibStab}
   \end{figure}

It is known that flare HXR emission exhibits a harder spectrum at
emission peaks than at valleys \citep[e.g.,][]{kiplinger83}. The
spectral analysis  for spikes agrees with this general scenario.
The different spectral characteristics between fast-varying HXR
spikes and underlying components are likely a result of specific
physical mechanisms to generate nonthermal emissions in such a way
that the spectral properties are closely related to the burst
timescales.

\subsection{Energy-dependent time delay}

We analyzed the time-delay properties of all the spikes detected
by different criteria. Four energy bands (25--40, 40--60, 60--100,
and 100--300 keV) were used in our analysis. For each spike, we
found the peak times $\tau(\epsilon)$ at different energies
relative to the peak time in 25--100 keV. We then performed a
least-squares linear-fit between the mean photon energy and
$\tau(\epsilon)$ in logarithmic scale for each spike. If the
fitting error was small,  we assumed  that the spike exhibits a
systematic energy-dependent lag in its peak times. Indeed, less
than 40\% of the spikes evolve with regular time delay patterns,
and both low-energy delay and high-energy delay events are
detected (note that in Paper I, we only found low-energy delay
events). In Figure 12, we illustrate the distribution of time lags
between 60--100 keV and 25--40 keV for these events. In the
figure, positive time lags indicate that high-energy emission lags
 behind low-energy emission, and negative time lags indicate that
low-energy emission lags behind high-energy emission. Evidently,
the majority of events exhibit time lags shorter than 0.5 s. The
mean time lag is about  0.80  s and  $-0.74$  s for high-energy
delays and low-energy delays, respectively.

We additionally investigated energy-dependent time lags with
respect to the spectral properties for a small fraction of spikes
that exhibit a systematic time lag pattern as well as reliable
spectral fitting results (\S3.2). For these spikes, we derived
spike peak times $\tau(\epsilon)$ at different energies relative
to a reference energy (25--100 keV), and fit $\tau (\epsilon)$ to
an arbitrary function $\tau (\epsilon) \sim \epsilon ^{\alpha}$. A
positive $\alpha$ indicates that peak emission in higher energies
lags behind the peak emission in lower energies, and a negative
$\alpha$ indicates lower energy emission lagging behind higher
energy emission. The time delay parameter $\alpha$ for different
populations of the spikes is shown in Figure 13.  We divided the
spikes into two types: low-energy delayed (negative $\alpha$) and
high-energy delayed (positive $\alpha$). Accordingly, the left
part of each panel in Figure 13 displays the spike spectral index
distribution for low-energy delayed events, while the right part
shows the distribution for high-energy delayed spikes. In the last
panel, the sample number is two, and both of them are low-energy
delayed spikes. These figures show that, among these events, about
or more than 2/3 are low-energy delayed, and the rest are
high-energy delayed. Furthermore, it is found that, on average,
high-energy delayed events have a harder count spectrum than
low-energy delayed events.

\begin{figure}
   \includegraphics[width=9.5cm]{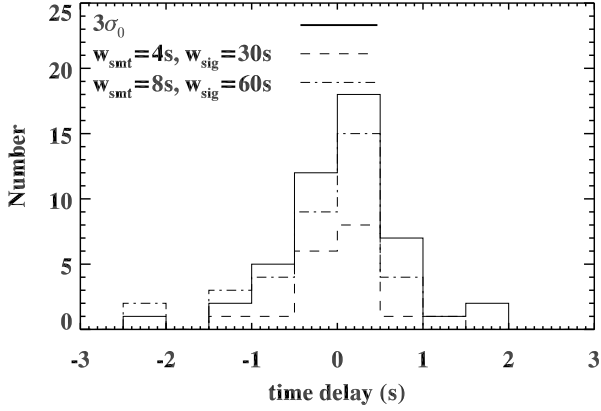}
      \caption{Distributions of spike peak time lags between 60--100 keV and 25--40 keV for spikes selected with various criteria.}
    \label{FigVibStab}
   \end{figure}

The energy-dependent time lag patterns associated with spike
energetics may result from the acceleration and/or transport
mechanisms. For example, depending on the initial pitch-angle
distribution, when a significant amount of electrons are trapped
in the corona, Coulomb collisions in the trap will eject these
electrons from the trap to precipitate and lose their energy in
the chromosphere. The rate of Coulomb collision is
energy-dependent, causing low-energy electrons to precipitate
first, thus producing high-energy delayed events. On the other
hand, when most electrons are directly precipitated in the
chromosphere, the energy-dependent time delays are associated with
the electron time of flight, i.e., higher energy electrons with a
higher velocity will precipitate ahead of low-energy electrons,
accounting for the low-energy delayed events. Our preliminary
result showing high-energy delayed events with a harder spectrum
indicates that the energetics of HXR spikes are probably coupled
with the transport effects, which are, in the first place,
governed by acceleration mechanisms determining the electron
pitch-angle distribution. A better understanding of the mechanisms
that generate these spikes should be obtained from the knowledge
of their spatial and magnetic properties.

\begin{figure}
   \includegraphics[width=8.5cm]{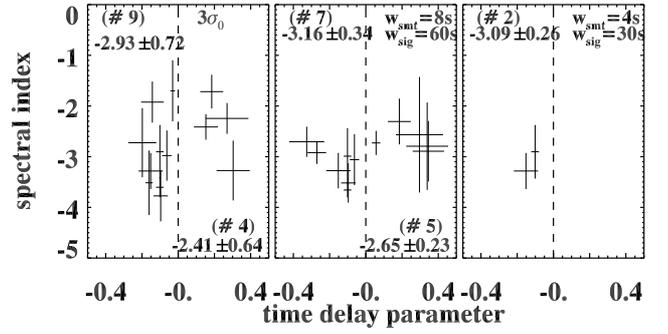}
      \caption{Scatter plot of the
spectral index versus the time delay parameter $\alpha$ of spikes
selected with various criteria. In each panel, positive $\alpha$
values refer to high-energy delayed events while negative ones
refer to low-energy delayed spikes. These two populations are
separated by the vertical dashed line, and the number of the
spikes, and the mean count spectral index $\Gamma$ and its
standard deviation of each population are also marked in each
panel. }
         \label{FigVibStab}
   \end{figure}
\section{Discussions and conclusions}

We have searched all flares observed by RHESSI in 2002 for
fast-varying HXR spikes, finding that at least 20\% of the flare
bursts with count rates exceeding 100 counts s$^{-1}$ in 25--100
keV produce HXR spikes, mostly during the rise phase of the
flares. These spikes have timescales of 0.2--2 s in photon energy
range of up to 300 keV. The main results are as follows. (1) Both
impulsive and long duration flares can produce HXR spikes with
nearly equal production rates. Flares with high peak count rates
are more productive in HXR spikes. (2) Almost all spikes occur in
the rise phase of the flares, and a large percentage, up to 70\%,
of spikes are produced at or about the flare peak times. (3) The
mean duration of spikes is about 0.9--1.0 s, independent of photon
energies. The rise and decay times of spikes are shown to be
almost the same. This differs from ordinary flares that usually
have a longer decay phase dominated by thermal emission. (4) Most
of the spikes can be detected in very high energy bands up to
100--300 keV. The HXR spectra of spikes are harder than those of
the underlying slow-varying components. This fact implies the
nonthermal origin of spikes. (5) Evident energy-dependent time
lags are present in a fraction of spikes, indicative of
time-of-flight or Coulomb collision effects. It is also shown
that, on average, spikes lagging in high-energy emissions have
harder spectra than spikes exhibiting lags in low-energy
emissions.

To understand the mechanisms for HXR spikes reported in this
study, more observational investigations are needed with
substantially improved capabilities or techniques to conduct
precise spectral and imaging analysis at the timescales of the
spikes.

\begin{acknowledgements}
We thank the referee for constructive comments that help improve
the presentation. We thank  G. J. Hurford for help with the
demodulation. Part of this work was conducted during the Research
Experience for Undergraduates (REU) program supported by National
Science Foundation through grant ATM-0552958 contracted to Montana
State University. This work was supported by the Scientific
Research Foundation of Graduate School of Nanjing University,
FANEDD under grant 200226, NSFC under grants 10878002, 10933003,
11133004 and 11103008, NKBRSF under grant 2011CB811402, the
Chinese Academy of Sciences (KZZD-EW-01-3), and US NASA grant
NNX08AE44G.

\end{acknowledgements}


\begin{thebibliography}{}
\bibitem[Aschwanden et al.(1995)]{asch95}Aschwanden, M. J., Schwartz, R. A., \& Alt, D. M. 1995, \apj,
447, 923

\bibitem[Aschwanden et al.(1997)]{aschwanden97}Aschwanden, M. J., Bynum, R. M., Kosugi, T., Hudson,
H. S., \& Schwartz, R. A. 1997, \apj, 487, 936

\bibitem[Aschwanden(2002)]{aschwanden02}Aschwanden, M. J. 2002, Particle
Acceleration and Kinematics in Solar Flares, A Synthesis of Recent
Observations and Theoretical Concepts, Kluwer Academic Publishers,
Dordrecht, Holland.

\bibitem[Aschwanden (2004)]{asch04}Aschwanden, M. J. 2004, Physics of the Solar Corona. An Introduction (Chichester, UK:
Praxis Publishing Ltd., Berlin: Springer-Verlag)

\bibitem[Bai \& Ramaty(1979)]{bai79}Bai, T., \& Ramaty, R. 1979,
\apj, 227, 1072

\bibitem[Bespalov, Zaitsev, \& Stepanov(1987)]{bespalov87}Bespalov, P. A., Zaitsev, V. V., \& Stepanov, A.
V. 1987, \solphys, 114, 127

\bibitem[Brown, Conway, \& Aschwanden(1998)]{brown98}
 Brown, J. C., Conway, A. J., \& Aschwanden, M. J. 1998, \apj, 509, 911

\bibitem[Brwon(1971)]{brown71}Brown, J. C. 1971, \solphys,
18, 489

 \bibitem[Carmichael(1964)]{car64} Carmichael, H. 1964, in Physics of Solar Flares,
ed. W. N. Hess (NASA SP-50; Washington DC: NASA), 451

\bibitem[Correia et al.(1995)]{correia95}Correia, E., Costa,
 J. E. R., Kaufmann, P., Magun, A., \& Herrmann, R. 1995, \solphys,
 159, 143

\bibitem[de Jager, \& de Jonger(1978)]{jager78}de Jager, C. \& de Jonge, G. 1978, \solphys, 58,
127

\bibitem[Emslie, \& Nagai(1985)]{emslie85} Emslie, A. G., \& Nagai,
F. 1985, \apj, 288, 779

\bibitem[Fisher, Canfield \& McClymont(1985)]{fisher85}Fisher, G. H., Canfield, R. C., \& McClymont, A.
N. 1985, \apj, 289, 414

\bibitem[Furth, Killeen, \& Rosenbluth(1963)]{furth63}Furth, H. P., Killeen, J.,
\& Rosenbluth, M. N. 1963, Phys. Fluids, 6, 459.

\bibitem[Hirayama(1974)]{hira74}Hirayama, T. 1974, \solphys, 34,
323

\bibitem[Hoyng, van Beek \& Brown(1976)]{hoyng76}Hoyng, P., van Beek, H. F. \& Brown, J. C. 1976, \solphys,
48, 197

\bibitem[Kaufmann(1996)]{kaufmann96}Kaufmann, P. 1996, \solphys,
169, 377

\bibitem[Kiplinger et al.(1983)]{kiplinger83}Kiplinger, A. L., Dennis, B. R.,
Frost, K. J., Orwig, L. E., \& Emslie, A. G. 1983, \apj, 265, L99

\bibitem[Kiplinger et al.(1984)]{kiplinger84}
Kiplinger, A. L., Dennis, B. R., Frost, K. J., \& Orwig, L. E.
1984, \apj, 287, L105

\bibitem[Kiplinger et al.(1989)]{kiplinger89}Kiplinger, A. L., Dennis, B. R., \& Orwig, L. E. 1989, Max '91
Workshop 2: Developments in Observations and Theory for Solar
Cycle 22 ed. R. M. Winglee \& B. R. Dennis (Greenbelt: NASA), 346

\bibitem[Kliem et al.(2000)]{kliem00}Kliem, B., Karlick, M., \& Benz, A. O. 2000,
\aap,  360, 715

\bibitem[Kopp \& Pneuman(1976)]{kopp76}Kopp, P. A., \& Pneuman,
G. W. 1976, \solphys, 50, 85

\bibitem[Lin et al.(2002)]{lin02}Lin, R. P., Dennis, B. R., Hurford, G. J., Smith, D. M.,
Zehnder, A., Harvey, P. R., Curtis, D. W., Pankow, D., Turin, P.,
Bester, M., and 56 coauthors 2002, \solphys, 210, 3

\bibitem[MacNeice et al.(1984)]{mac84}MacNeice, P., Burgess, A., McWhirter, R. W. P., \& Spicer, D.
S. 1984, \solphys, 90, 357

\bibitem[Mariska, Emslie, \& Li(1989) ]{mar89}
 Mariska, John T., Emslie, A. Gordon, \& Li, Peng 1989, \apj, 341, 1067

\bibitem[Melrose \& Brown(1976)]{mel76}Melrose, D. B. \& Brown, J.
C. 1976, MNRAS, 176, 15

\bibitem[Priest \& Forbes(2000)]{priest00}Priest, E., \& Forbes, T. 2000, Magnetic Reconnection
(Cambridge, UK: Cambridge University Press)

\bibitem[Qiu, Lee, \& Gary(2004)]{qiu04}Qiu, J., Lee, Jeongwoo, \& Gary, Dale E.  2004, \apj, 603, 335

\bibitem[Qiu et al. (2012)]{qiu12}
Qiu, J., Cheng, J. X., Hurford, G. J., Xu, Y., \& Wang, H. 2012,
\aap, accepted, Paper I

\bibitem[Sturrock(1966)]{Stu66}Sturrock, P. A., 1966, \nat, 211,
695

\bibitem[Sturrock \& Uchida(1981)]{sturrock81}Sturrock, P. A.,
\& Uchida, Y. 1981, \apj, 246, 331

\bibitem[Sturrock et al.(1984)]{sturrock84}Sturrock, P. A., Kaufman, P., Moore, R. L. \& Smith, D. F. 1984,
\solphys, 94, 341

\bibitem[van Beek, de Feiter \& de Jager(1974)]{van74}van Beek, H. F., de Feiter, L. D., \& de Jager, C. 1974, In:
Space research XIV; Proceedings of the Sixteenth Plenary Meeting,
Berlin, East Germany, Akademie-Verlag GmbH, p. 447-452.

\bibitem[van Beek, de Feiter \& de Jager(1976)]{van76}van Beek, H.
F., de Feiter, L. D., \& de Jager, C. 1976, In: Space research
XVI; Proceedings of the Open Meetings of Working Groups on
Physical Sciences, Berlin, East Germany, Akademie- Verlag GmbH, p.
819-822.

\bibitem[Vilmer, Kane \& Trottet(1982)]{vil82}Vilmer, N., Kane, S. R., \& Trottet, G. 1982, \aap, 108, 306
\end{thebibliography}
\end{document}